\documentclass[prb,amsmath,amssymb,floatfix,twocolumn,citeautoscript]{revtex4}
\usepackage{bm}
\usepackage{epsfig}

\begin{document}

\title{Non-contact Friction and Relaxational Dynamics of Surface Defects}

\author{Jian-Huang She$^1$, and Alexander V. Balatsky$^{1,2}$}

\affiliation{$^1$Theoretical Division, Los Alamos National Laboratory, Los Alamos, NM, 87545, USA.\\
$^2$Center for Integrated Nanotechnologies, Los Alamos National Laboratory, Los Alamos, NM, 87545, USA.}

\begin{abstract}
Motion of cantilever near sample surfaces exhibits additional friction even before two bodies come into mechanical contact. Called non-contact friction (NCF),  this friction is of great practical importance to the ultrasensitive force detection measurements. Observed large NCF of a  micron-scale cantilever found anomalously large damping that exceeds theoretical predictions by 8-11 orders of magnitude. This finding points to contribution beyond fluctuating electromagnetic fields within van der Waals approach.  Recent experiments reported by Saitoh et al.  (Phys. Rev. Lett. 105, 236103 (2010)) also found nontrivial distance dependence of NCF. Motivated by these observations, we propose a mechanism based on the coupling of cantilever to the relaxation dynamics of surface defects. We assume that the surface defects couple to the cantilever tip via spin-spin coupling and their spin  relaxation dynamics gives rise to the backaction terms and modifies both the friction coefficient and the spring constant.  We  explain the magnitude, as well as the distance dependence of the friction due to these backaction terms. Reasonable agreement is found with the experiments.
\end{abstract}

\date{\today \ [file: \jobname]}

 \maketitle

\begin{figure}
\begin{centering}
\includegraphics[width=0.6\linewidth]{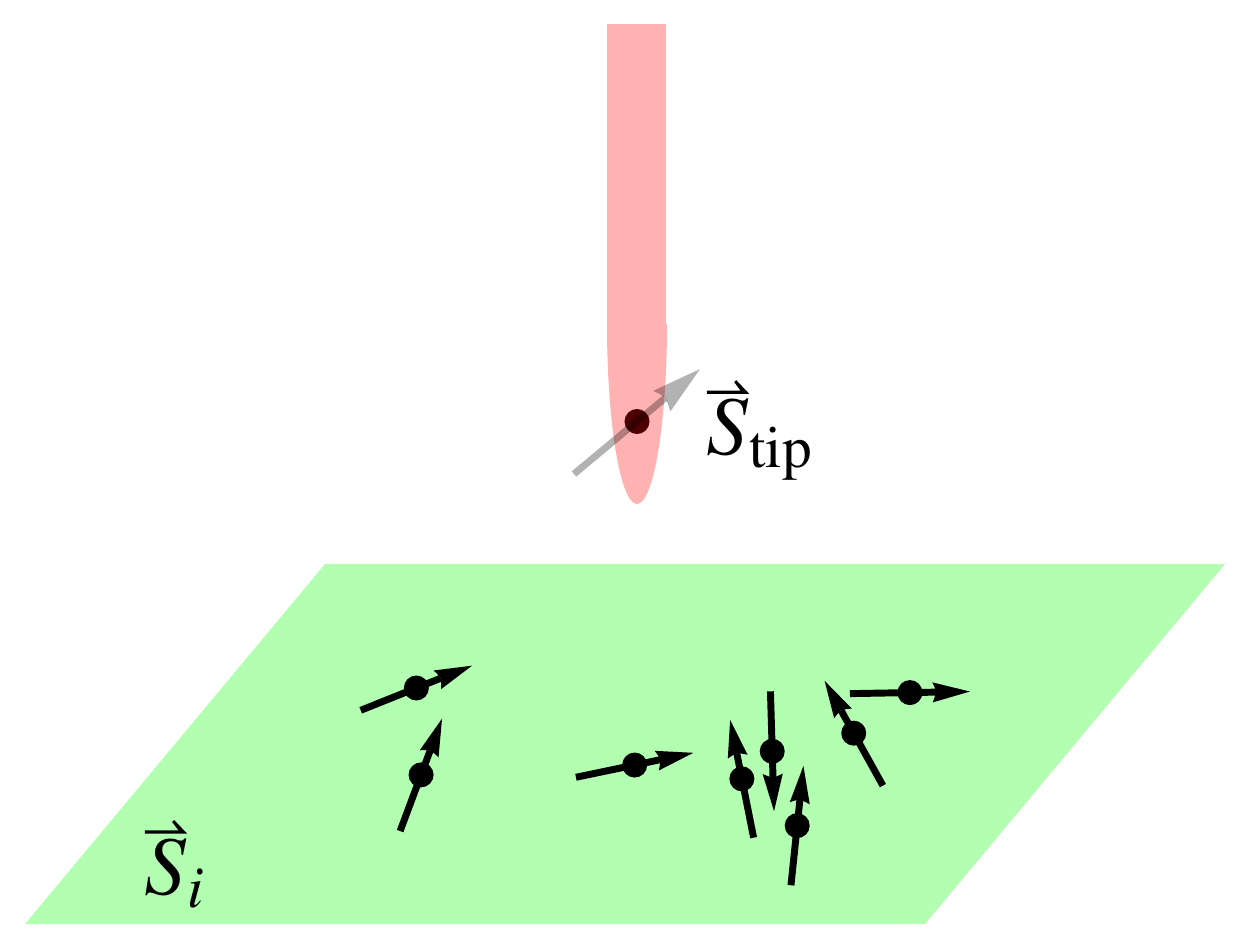}
\end{centering}
\caption{(Color online) Illustration of proposed mechanism that gives rise to the non-contact friction between the cantilever tip and the sample surface: randomly distributed defect spins on the sample surface interacting with the spins residing on the tip.}
\end{figure}

{\it Introduction:} Friction is one of the most widely perceived but least understood phenomena in nature. Friction is ubiquitously seen around us at the macroscale. One might expect however that new insights will be gained by investigating friction mechanisms at small scale and short distances. Indeed, recent advances in nanotechnology have enabled the study of friction on the nanoscale \cite{Bhushan}, where a novel form of friction has been discovered, namely the non-contact friction (NCF) \cite{Dorofeyev99, Stipe01, Kuehn06}. This kind of friction occurs when two objects are in close proximity but not in physical contact. NCF is of great practical importance for the modern development of ultrasensitive force detection devices \cite{Sidles, Giessibl}. The precision of these measurements may be ultimately limited by the effects of NCF.

The origin of NCF is still under debate. One proposal was that it is the friction resulting from Ohmic losses mediated by fluctuating electromagnetic fields. It turns out be 8-11 orders of magnitude smaller than that observed in experiments \cite{Dorofeyev99, Stipe01, Persson02, Chumak04}. Several alternative mechanisms have been proposed in recent years (see \cite{Persson07} and references therein), but the problem remains unsolved. More recently, systematic studies of NCF between a cantilever tip and the sample surface have been performed on metallic, insulating and superconducting materials at different temperatures using hard cantilevers \cite{Saitoh10} (as opposed to soft cantelevers results \cite{Stipe01} ). These studies found that at low temperatures, the friction coefficient caused by a superconducting sample is an order of magnitude larger than that of an insulating sample, in contradiction with the previous theoretical prediction that NCF generically scales with the resistivity of the sample \cite{Persson07}. Furthermore, a universal feature has been identified in these experiments \cite{Saitoh10}, namely, at low temperatures, while the induced spring constant increases monotonically with decreasing distance, the friction coefficient displays a maximum at certain distance of a few nanometers. Such feature has been consistently observed in insulating materials as well as superconducting materials, both below and above the superconducting transition temperature.

In this paper, we propose an explanation for these experimental findings \cite{Saitoh10}. Specifically, we propose that the observed  behavior of NCF is due to the relaxation dynamics of the surface defects.

 i) First we lay out the general formalism. It is illustrated by considering localized spins.  We assume that spins remain unscreened due to the insulating or  superconducting gap. We assume that cantilever also carries localized spins that couple to the collection of spin sites on the surface of the sample which have characteristic relaxation time $\tau_d$ as function of tip-sample distance $d$. We model this relaxation time as a collection of Debye relaxors with distributed relaxation times that results in a typical glass-like backaction dynamics, see below. Analysis of the experimental data indeed indicates a distribution of relaxation times \cite{Saitoh10}. Alternatively defects can be local charged sites. In that case the charging and uncharging dynamics of the defects leads to the random electric field probed by a cantilever.

 ii)Using our general formalism we then extract from experimental data the distance dependence of tip-sample coupling $A(d)$ and relaxation time $\tau_d$. $A(d)$ can be fitted by simple power laws, but $\tau_d$ displays surprising behavior. For extremely hard cantilever at low temperature, clear divergence in $\tau_d$ is observed. Weaker singularities are seen for moderately hard cantilevers. We propose an explanation for such behavior.

  iii)  We show next that the combined effect of the variation of  the tip-sample coupling and the relaxational dynamics on the sample surface explains the observed distance dependence of the friction coefficient and the induced spring constant. We also estimate the order of magnitude of the friction coefficient, and the experimentally observed values can be easily achieved in our framework.

{\it Formalism:} We model the tip of the cantilever by a massive particle with effective mass $m$, moving in a one-dimensional harmonic potential $V=\frac{1}{2}kx^2$. Here $k$ is the spring constant and $x$ the displacement. We consider that there are some randomly distributed active degrees of freedom on the sample surface that interact with the tip. Generically such interaction will then produce a frictional force, damping the motion of the tip. The tip motion is thus governed by the generalized Langevin equation
\begin{equation}
m\frac{d^2}{dt^2}x(t)+kx(t)+\int_{t_0}^{t} \gamma(t-t')
\frac{d}{dt'}x(t')dt'=F_x(t),
\label{GLE}
\end{equation}
with $\gamma(t)$ the backaction term that we write as a dynamical damping. There is also a residual random force $F_x(t)$, which fluctuates rapidly. For an equilibrium system, the frictional force and the random force are connected by the fluctuation-dissipation theorem (FDT) \cite{Kubo,Kubo66},
\begin{equation}
\gamma(\omega)=\frac{1}{k_BT}\int_0^{\infty}dte^{-i\omega t}\langle F_x(t_0)F_x(t_0+t) \rangle,
\label{FDT2}
\end{equation}
where $\gamma(\omega)$ is the Fourier transform of $\gamma(t)$, i.e.,$\gamma(\omega)=\int_0^{\infty}dte^{-i\omega t}\gamma(t)$. $\gamma(\omega)$ is generally a complex function. One can write it as $\gamma(\omega)\equiv\Gamma(\omega)-\frac{i}{\omega}k_{\rm int}(\omega)$. The real part $\Gamma$ describes the effect of dissipation, and the imaginary part leads to a modification of the spring constant.  $\Gamma=\Gamma_0+\Gamma_{\rm int}$, with $\Gamma_0$ the intrinsic cantilever friction and $\Gamma_{\rm int}$ the NCF resulting from tip-sample interaction.

{\it Spin-spin interaction:} The mechanism we consider is quite general. It applies to interactions in different channels, e.g. spin or charge. To be concrete, we will study in detail the spin-spin interaction.
 Since essentially the same relaxational behavior was observed in superconductors (NbSe$_2$) and insulators (SrTiO$_3$) \cite{Saitoh10}, spin-spin interaction is also a plausible choice. We notice that spin relaxation of surface magnetic defects is regarded as the origin of 1/f flux noise in superconducting devices \cite{Koch07, Sousa07, Sendelbach08}.

 We consider there are randomly distributed localized spins on the sample surface and they interact with the spin localized on the tip, with a Hamiltonian $H=\sum_{ia}J_i^a(x)S^a_{\rm tip}S^a_i$. Here $S^a_{\rm tip}$ is the spin operator on the tip, $S^a_i$ the spin operator on the sample surface, and the coupling $J_i^a$ can be of different types, e.g. Ising, Heisenberg.
The force in the $x$ direction is $F_x=-\partial H/\partial x$. $\gamma(\omega)$ thus reads
\begin{align}
\gamma(\omega)=&\frac{1}{k_BT}\lim_{x,x'\to0}\int_0^{\infty}dte^{i\omega t}\frac{\partial}{\partial x}\frac{\partial}{\partial x'}\nonumber\\
&\sum_{ijab}\langle J^a_i(x)J^b_j(x') S^a_{\rm tip}(0)S^a_i(0)S^b_{\rm tip}(t)S^b_j(t) \rangle.
\end{align}

The spins on the sample surface interact with each other, and the tip spin provides an external magnetic field ${\mathbf H}$ for the surface spin system. The Hamiltonian of the surface spin system is thus of the form, ${\cal H}_{\rm surf}=-\sum_{i\neq j}J_{ij}^aS^a_i S^a_j-\sum_{i}H^a_iS^a_i$.
Effect of the external field on dynamics of spins is that it will change the relaxation time $\tau$, see below. The surface spin system displays disordered behavior, where the cross correlations vanish and the dynamics is characterized by the autocorrelation function, with $\langle S^a_i(t) S^a_j(0) \rangle=q(t)\delta_{ij}$.

We treat the dynamics of the spins on the sample surface is much faster than that of the tip spin due to the significantly higher density of scattering centers at the surface, hence cantilever spin can be viewed as static.
$\gamma(\omega)$ can then be factorized into two parts,
\begin{equation}
\gamma(\omega)=\frac{C_s}{k_BT}{\cal A}(d) {\cal S}_d(\omega).
\end{equation}
The prefactor $C_s$ comes from spin degeneracy.
The frequency-independent tip-sample coupling ${\cal A}(d)=\lim_{x\to0}\langle\sum_{i}\left(\frac{\partial}{\partial x}J^a_i(x)\right)^2\rangle$, increasing monotonically with decreasing $d$.  Introducing a defect density $\rho(\mathbf r)=\sum_{i}\delta({\mathbf r}-{\mathbf R}_i)$, which has average value $\langle\rho\rangle=n_{imp}$, one obtains
${\cal A}(d)=\langle\int d{\mathbf r} \rho({\mathbf r})J'({\mathbf r})^2\rangle=n_{imp}\int d{\mathbf r}J'({\mathbf r})^2$.
Here we have defined $J'({\mathbf r})=\sum_i\lim_{x\to0}\frac{\partial}{\partial x}J_i(x)\delta({\mathbf r}-{\mathbf R}_i)$.

The surface spin susceptibility ${\cal S}_d(\omega)=\int_0^{\infty}dt e^{-i\omega t}\langle S^a_i(0)S^a_i(t) \rangle$ is related to the response function $C(i\omega_n)=\int_0^{\beta} d\tau e^{-i\omega_n t}\langle S^a_i(0)S^a_i(t) \rangle$ by the classical form of the fluctuation-dissipation theorem ${\Re}{\cal S}(\omega)=(k_BT/\omega){\Im} C(\omega)$, or ${\cal S}(\omega)=-i (k_BT/\omega)C(\omega)$.

\begin{figure*}
\begin{centering}
\includegraphics[width=0.23\linewidth]{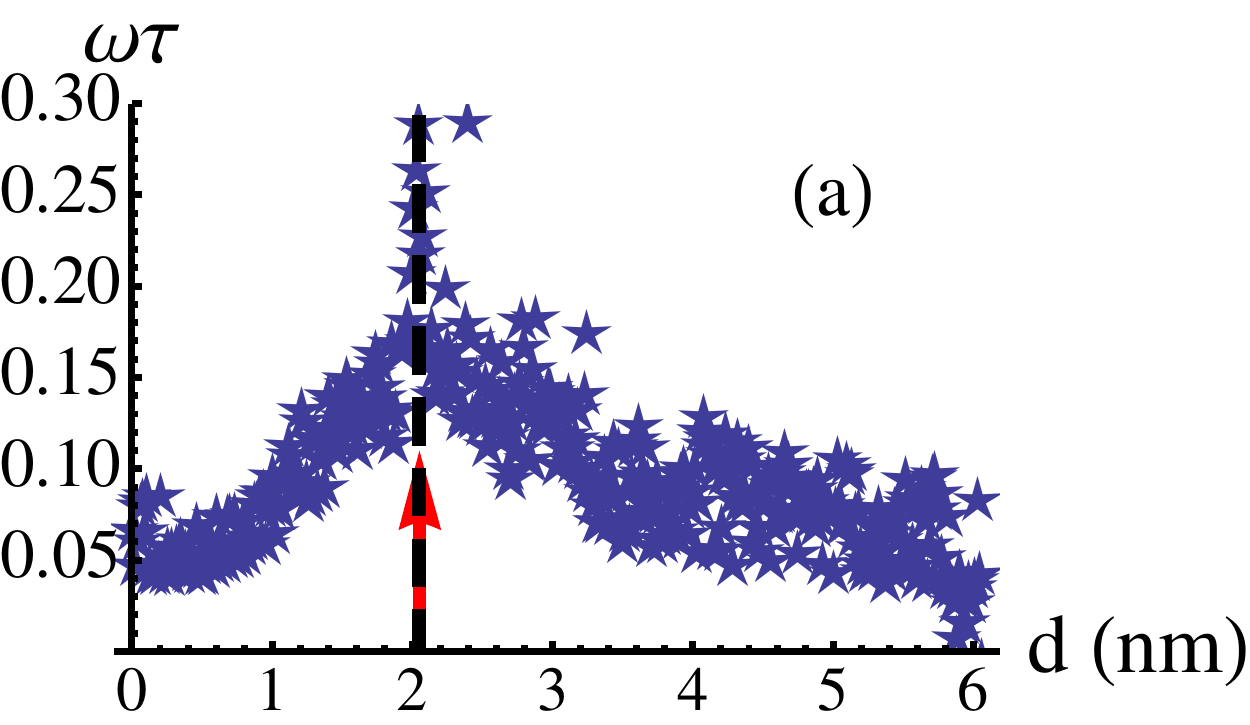}
\includegraphics[width=0.23\linewidth]{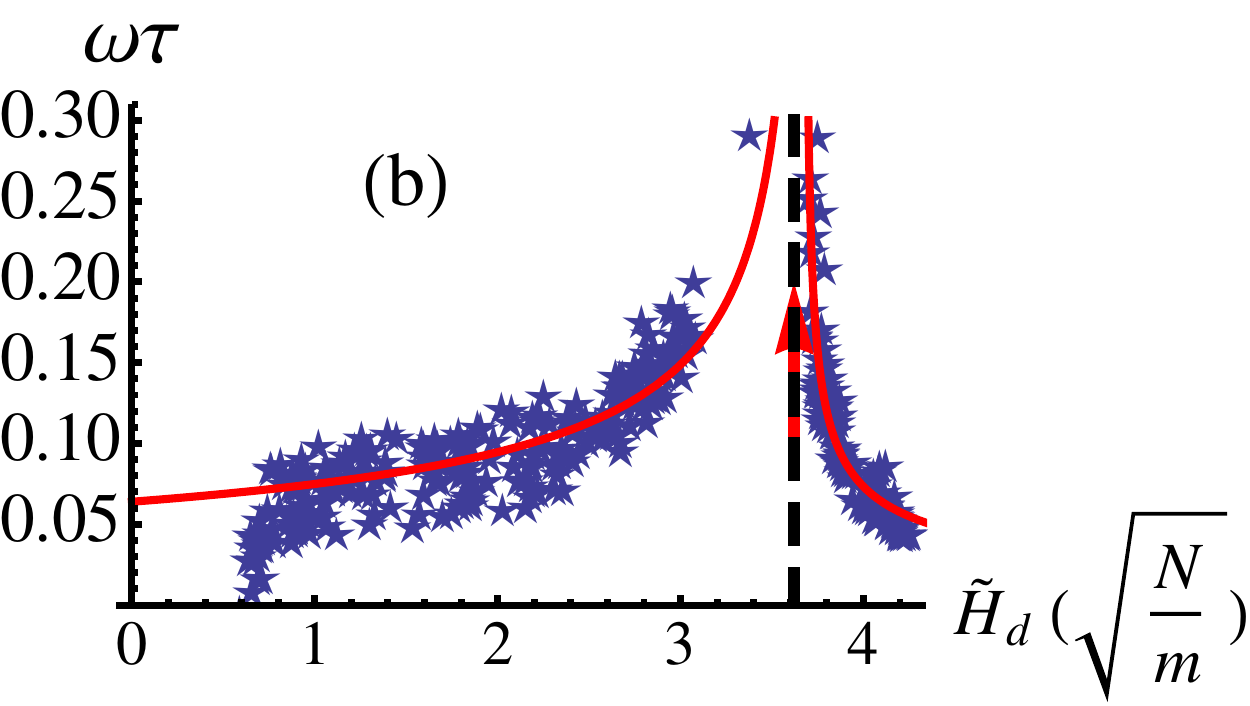}
\includegraphics[width=0.23\linewidth]{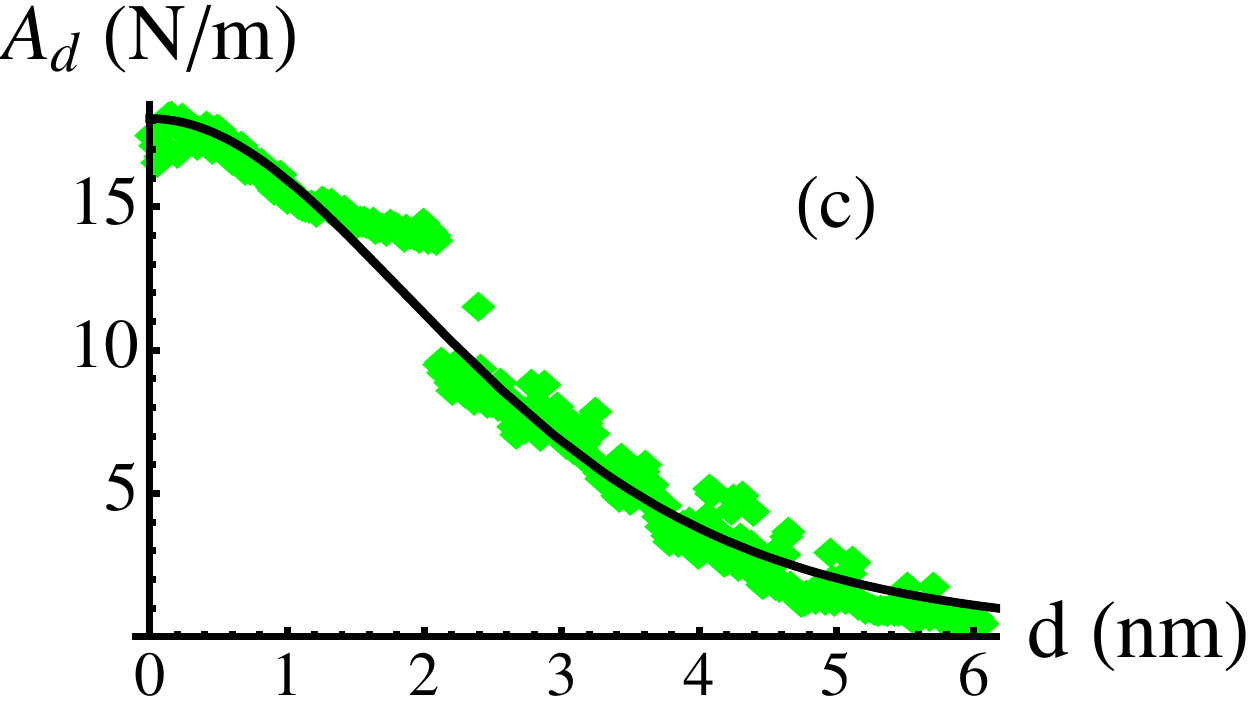}
\includegraphics[width=0.23\linewidth]{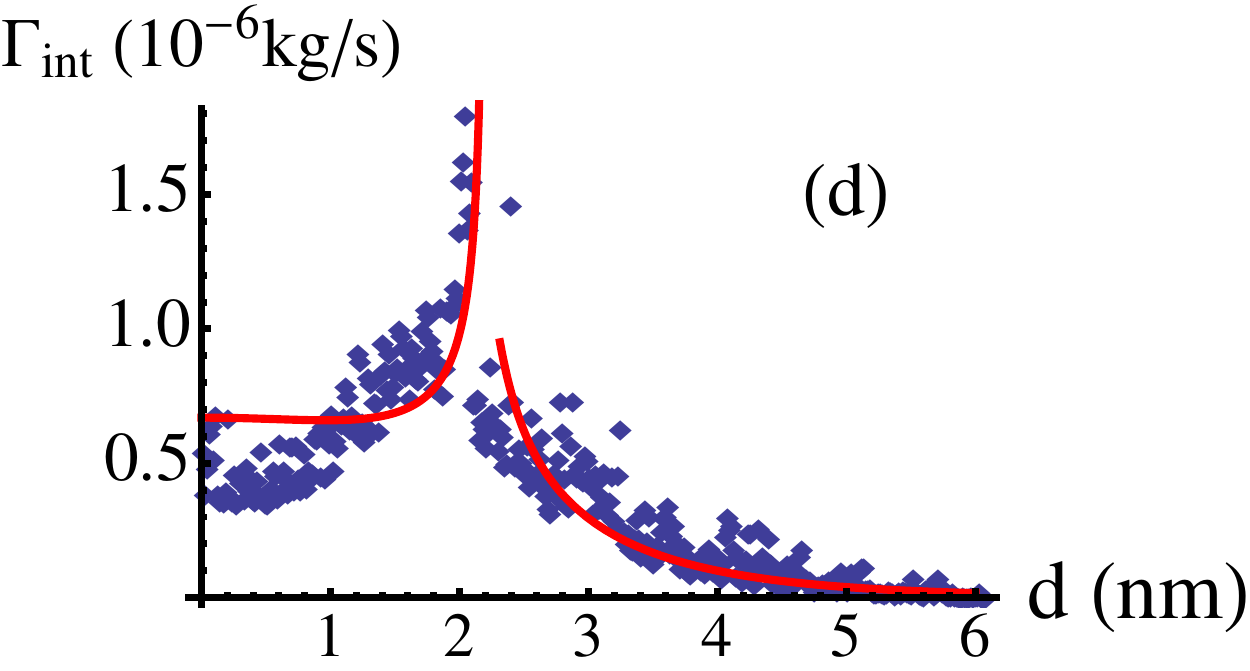}
\includegraphics[width=0.23\linewidth]{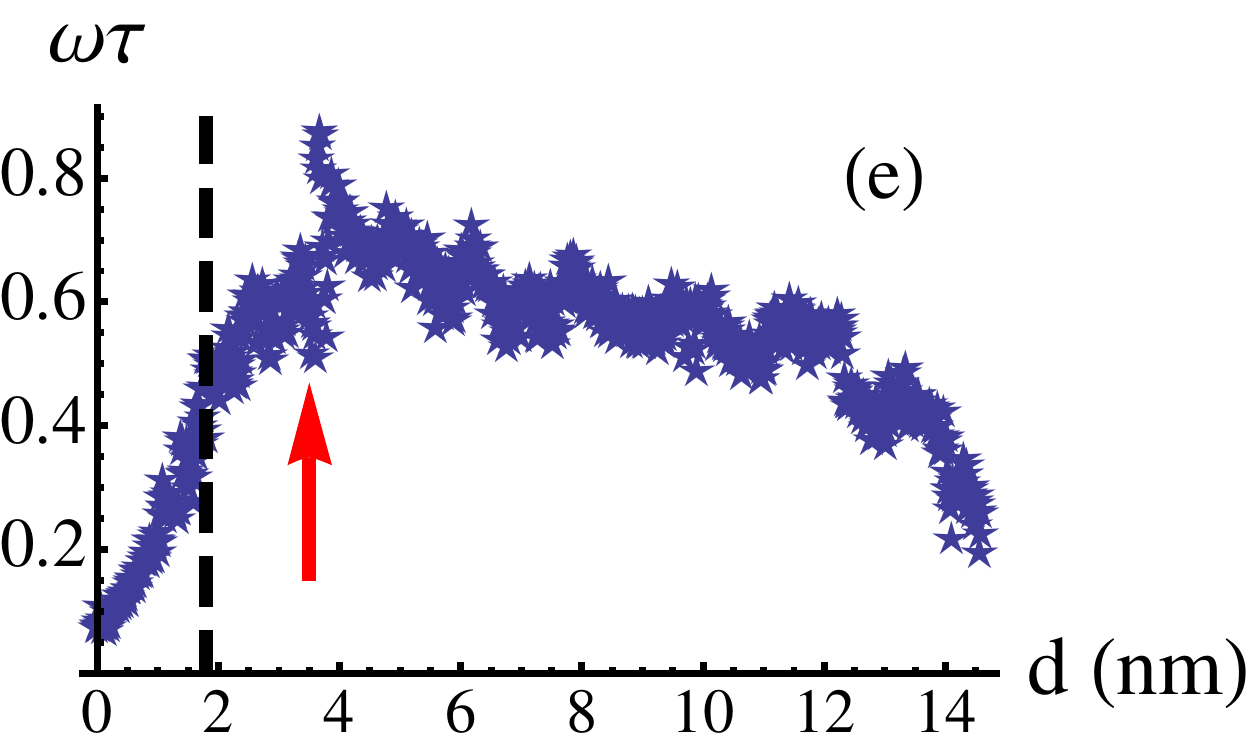}
\includegraphics[width=0.23\linewidth]{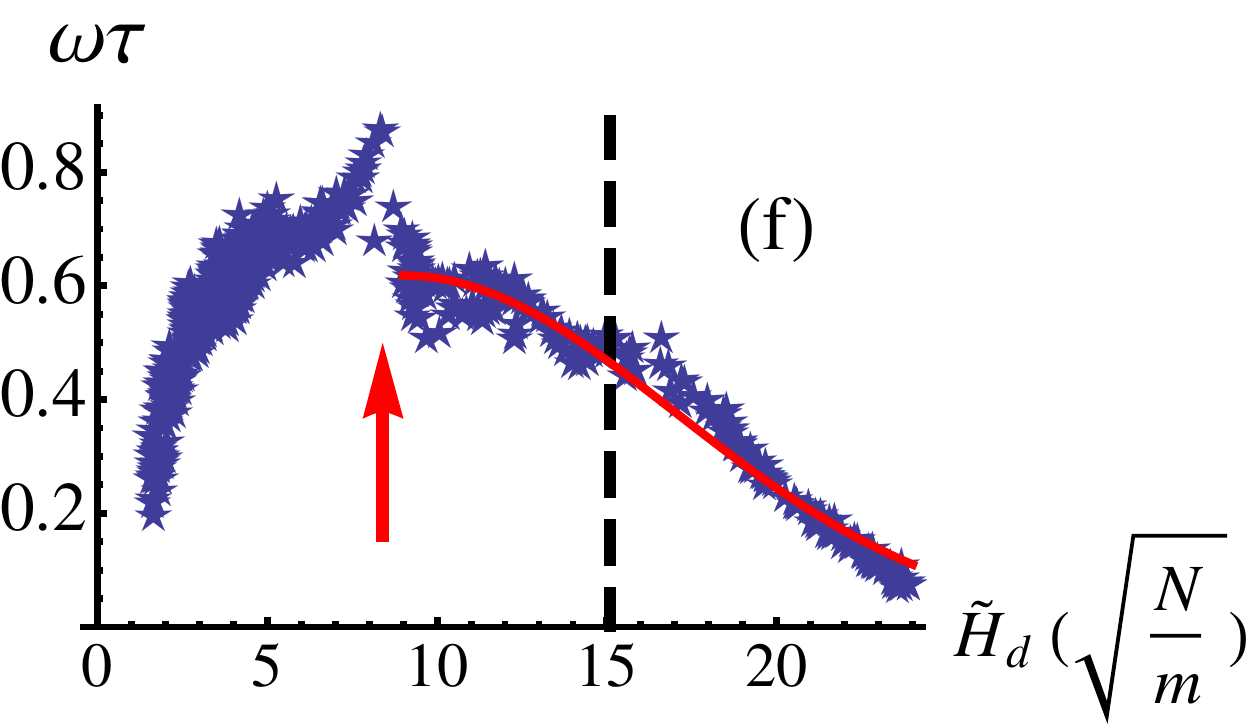}
\includegraphics[width=0.23\linewidth]{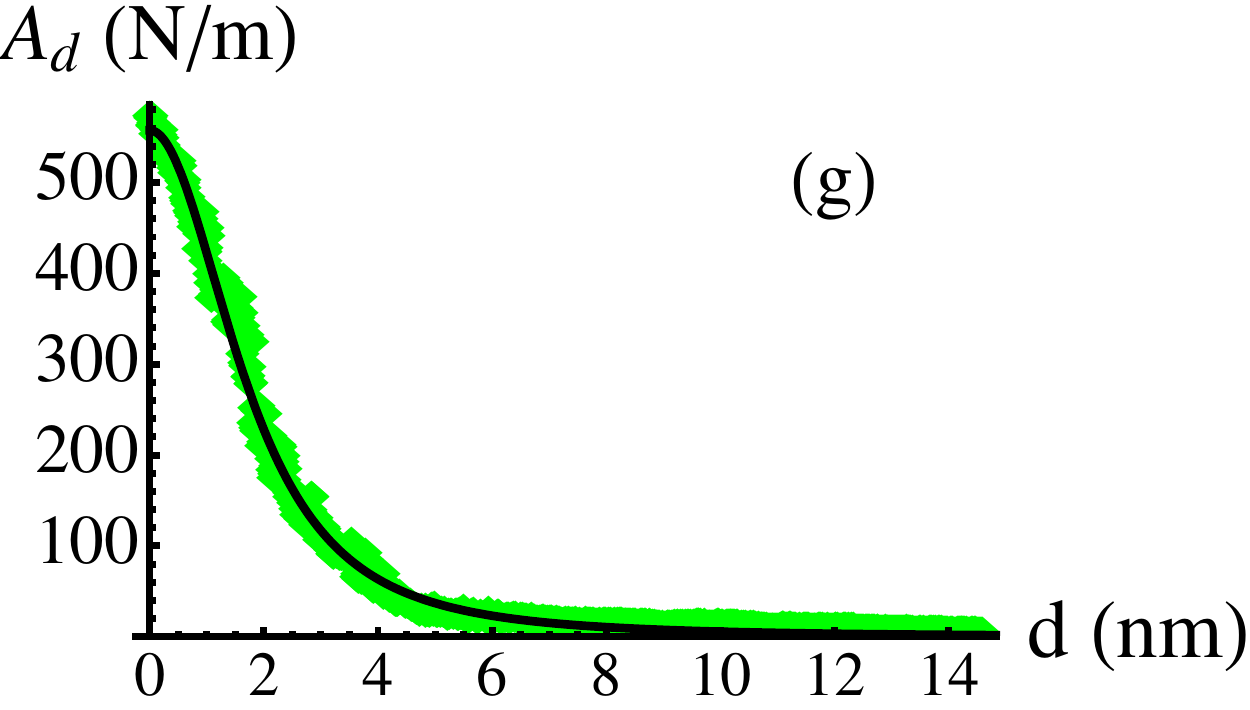}
\includegraphics[width=0.23\linewidth]{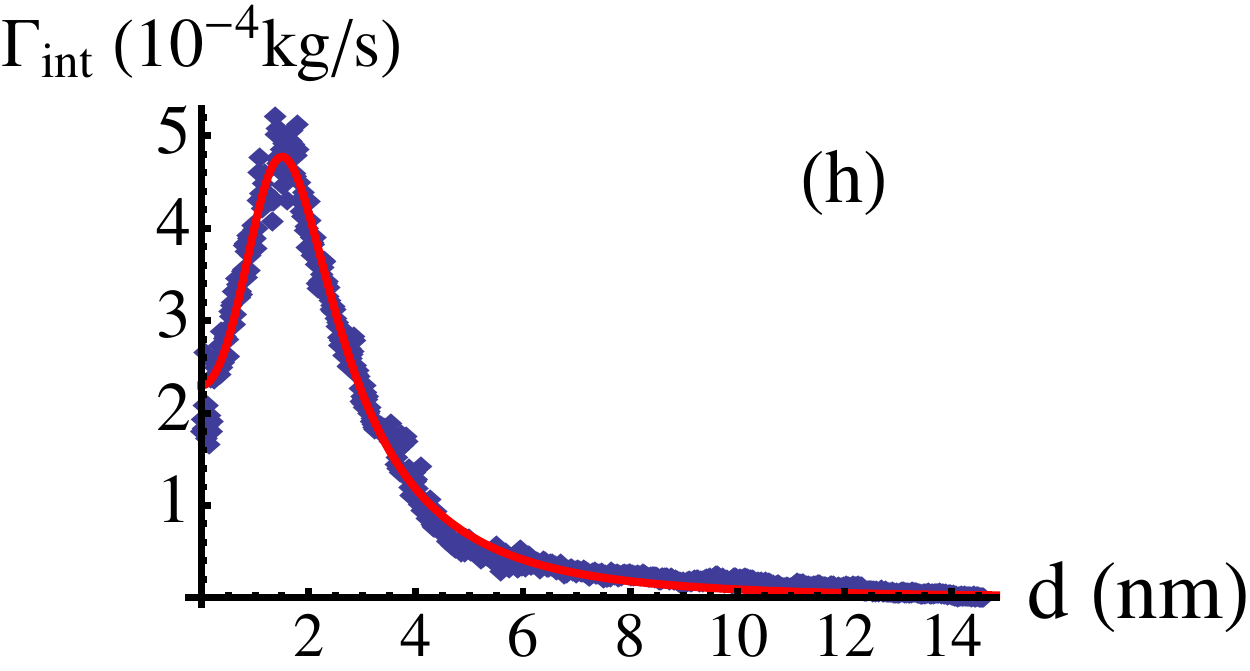}
\includegraphics[width=0.23\linewidth]{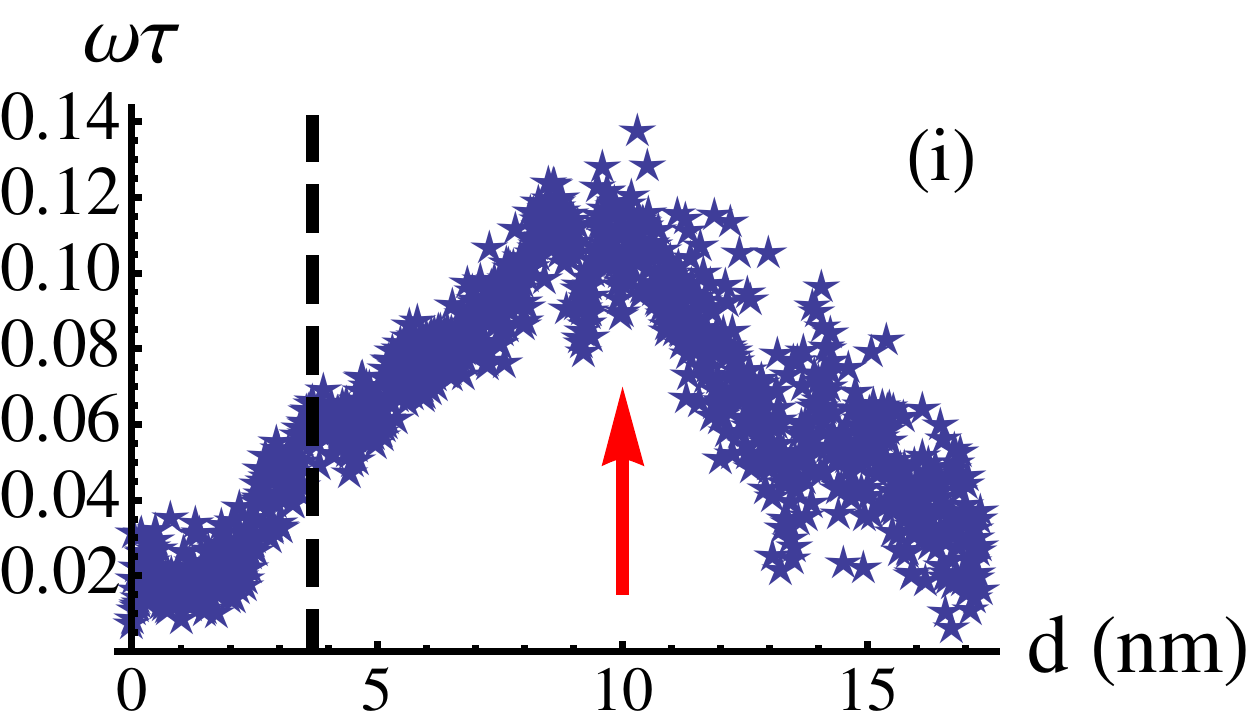}
\includegraphics[width=0.23\linewidth]{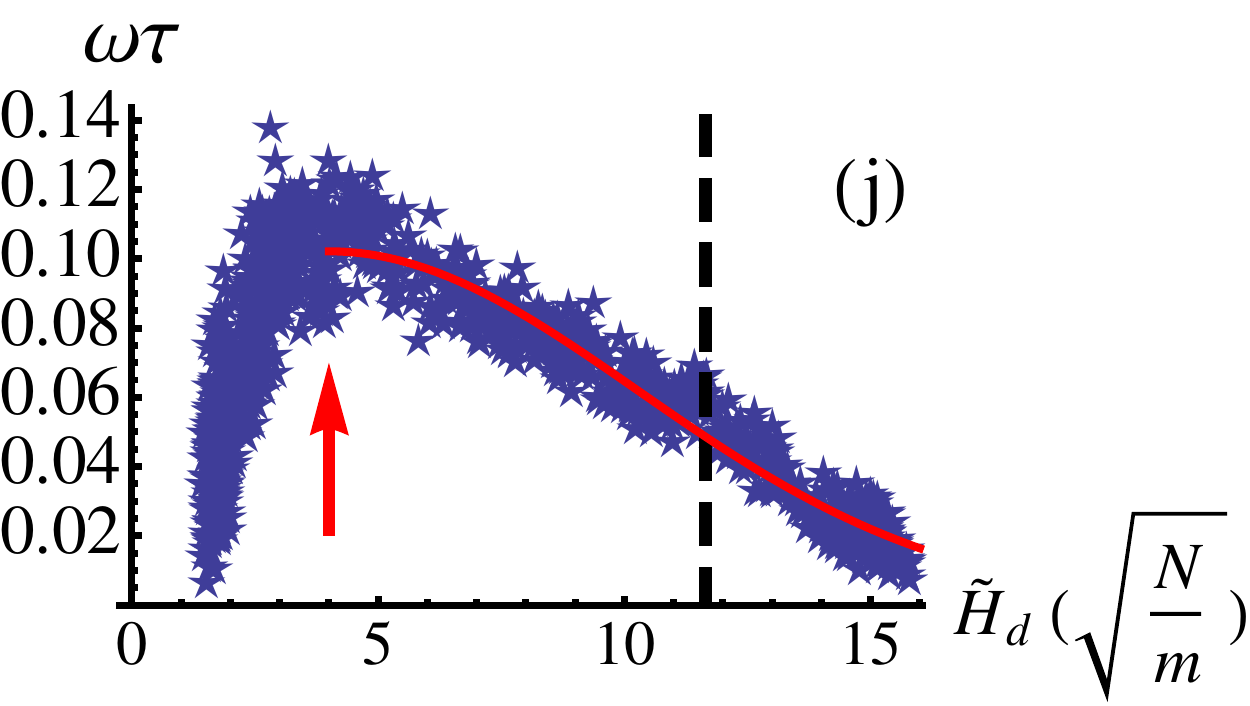}
\includegraphics[width=0.23\linewidth]{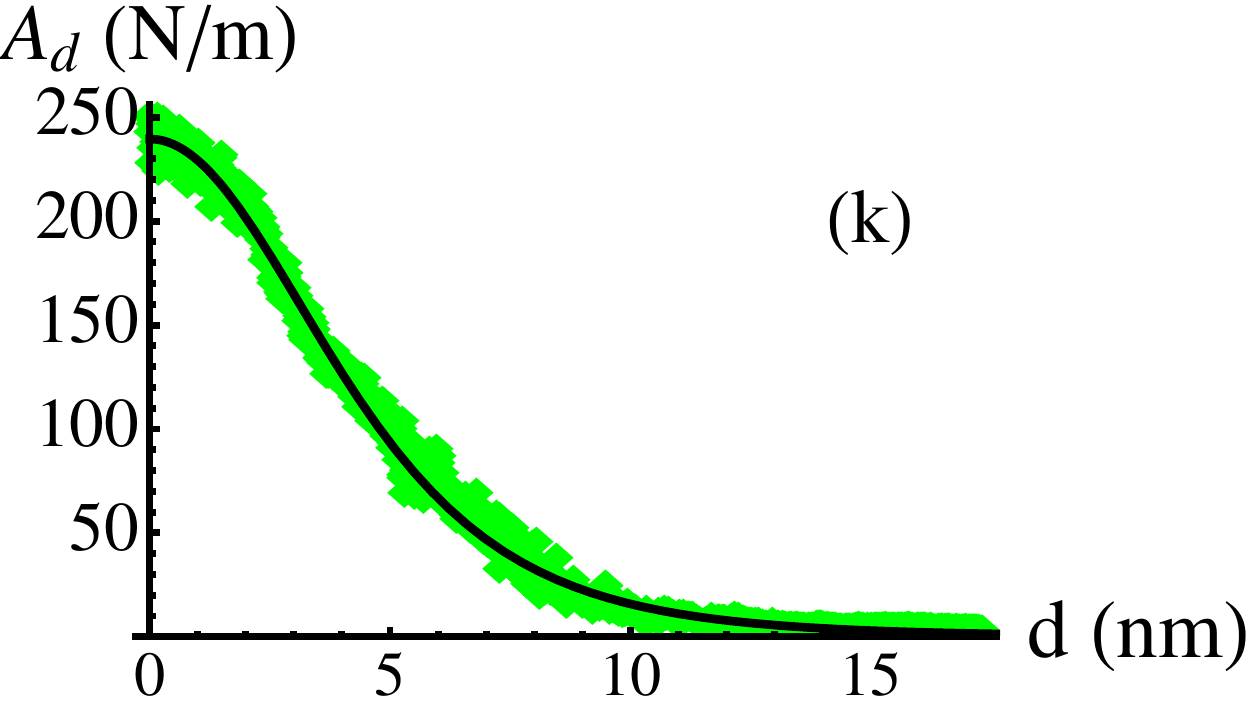}
\includegraphics[width=0.23\linewidth]{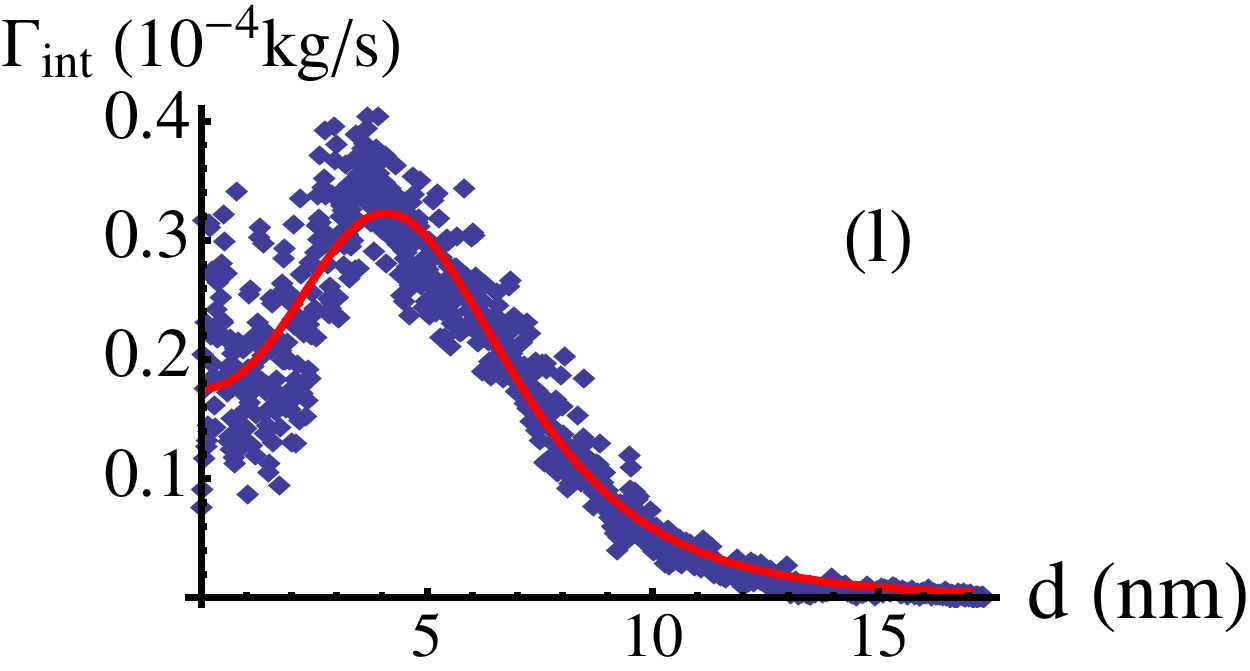}
\end{centering}
\caption{(Color online) Relaxation time, tip-sample coupling and friction at low temperature $T=4.2K$ for i) NbSe$_2$ and $f_0=300 {\rm kHz}$, ii) NbSe$_2$ and $f_0=31.6 {\rm kHz}$, iii) SrTiO$_3$ and $f_0=31.6 {\rm kHz}$, from top to bottom. The star and diamond are experimental values (data by Saitoh et al. \cite{Saitoh10} ). The dashed lines denote where $\Gamma_{\rm int}$ has a maximum. We use Eq.\ref{friction1} to fit $\Gamma_{\rm int}$. The glass exponent is chosen to be $a=0.9$. Since $H_d\propto \sqrt{A_d}$, we use $A_d$ as a measure of the magnetic field $H_d$, and define ${\tilde H}_d\equiv\sqrt{A_d}$.  For case i), $\omega\tau_d=C_{L,R} \left|{\tilde H}_d-H_{\rm SP}\right|^{-1/2}$, with $C_{L}=0.12, C_R=0.04$.  For cases ii), iii), $\omega\tau_d=C \exp(-({\tilde H}_d-H_{\rm SP})^2/H_0^2)$ with $C^{(ii)}=0.6,H_{\rm SP}^{(ii)}=9, H_0^{(ii)}=11.4$ and $C^{(iii)}=0.1,H_{\rm SP}^{(iii)}=4, H_0^{(iii)}=8.9$. The tip-sample coupling $A(d)=A_0(d^2+r_0^2)^{-\alpha}$, with $\alpha^{(i)}=3, A_0^{(i)}=2.3\times 10^5, r_0^{(i)}=4.8$; $\alpha^{(ii)}=1.5, A_0^{(ii)}=6.0\times 10^3, r_0^{(ii)}=2.2$; and $\alpha^{(iii)}=3, A_0^{(iii)}=7.5\times 10^7, r_0^{(iii)}=8.2$. We notice that for case i) there is a jump in $A_d$, which is "inherited" from the singularity in $\omega\tau_d$, and would disappear in a more realistic modeling. The high temperature result is included in Appendix II.}
\label{fit4K}
\end{figure*}

The interactions among the surface spins are random, leading to glassy behavior. Experiments also indicate a distribution of relaxation times \cite{Saitoh10}.
We thus assume $C(\omega)$ to have the usual phenomenological form typical for a glass system
$C(\omega)=C_0/(1-i\omega\tau_d)^a$,
with the exponent $0<a\leq 1$, and the relaxation time $\tau_d$ \cite{Cole41,Cole50,Cole51}.
$\gamma(\omega)$ now takes the form
\begin{equation}
\gamma(\omega)=-i\frac{C_1}{\omega}{\cal A}(d) \frac{1}{(1-i\omega\tau_d)^a},
\end{equation}
For cantilevers with high quality factor \cite{Saitoh10}, the intrinsic cantilever friction can be ignored. Defining $A(d)=C_1 {\cal A}(d)$, with $C_1=C_sC_0$, the NCF thus reads
\begin{equation}
\omega\Gamma_{\rm int}(\omega)= A(d)\frac{\sin[a\arctan (\omega\tau_d)]}{(1+(\omega\tau_d)^2)^{a/2}},
\label{friction1}
\end{equation}
and the induced spring constant is
\begin{equation}
k_{\rm int}(\omega)=A(d)\frac{\cos[a\arctan (\omega\tau_d)]}{(1+(\omega\tau_d)^2)^{a/2}}.
\label{spring1}
\end{equation}
Their ratio
\begin{equation}
\omega\Gamma_{\rm int}/k_{\rm int}=\tan[a\arctan (\omega\tau_d)]
\label{ratio1}
\end{equation}
 depends only on the relaxation dynamics of surface defects. We first extract $\omega\tau_d$ from the experimental data using Eq.\ref{ratio1}, and then calculate the tip-sample coupling $A(d)$ from Eq.\ref{friction1} or Eq. \ref{spring1}. The results are shown in Fig. \ref{fit4K}.

We approximate the effect of the tip spin as producing an uniform magnetic field $H_d$ on the sample surface.
When the spin-spin interaction decays with distance as $J(l)\sim l^{-\alpha}$, one has $A(d)\sim (d^2+r_0^2)^{-\alpha}$, and $H_d\sim (d^2+r_0^2)^{\alpha/2}$, with $r_0$ a cutoff. If $J(l)\sim \exp(-l^2/r_A^2)$, then $A(d)\sim \exp(-2d^2/r_A^2)$ and $H_d \sim \exp(-d^2/r_A^2)$. $A(d)$ can be fit with simple power laws (see Fig. \ref{fit4K}(c)(g)(k)).

Noticing $H_d\sim \sqrt{A(d)}$, we also extract the field dependence of relaxation time in Fig. \ref{fit4K}(b)(f)(j). The most surprising result is that for NbSe$_2$ probed by extremely hard cantilever (resonance frequency $f_0=300{\rm kHz}$), the relaxation time shows clear divergence when approaching certain field strength. This indicates that the surface spin system falls into the mean field university class and is consistent with behavior represented by the long-range, weakly interacting Husimi-Temperley model (see Appendix I). In this model, as one increases the magnetic field, the free energy changes from a double well structure to a single well structure, and the metastable minimum and the barrier combine at certain value of field strength to form a saddle point, known as the spinodal point. Near the spinodal point $H=H_{\rm SP}$, the relaxation of the system slows down dramatically, $\tau\sim\left|H-H_{\rm SP}\right|^{-1/2}$  \cite{Binder73,Mori10}. We use these predictions to model spin behavior here. The experimental result of $\omega\tau$ for NbSe$_2$ with $f_0=300{\rm kHz}$ can be fitted by such mean field form (see Fig. \ref{fit4K}(b)). In models with shorter range interactions, the change in relaxation time is smeared \cite{McDonald62, McDonald63, Rikvold94}, as can be seen in Fig. \ref{fit4K}(f)(j) for the two cases with $f_0=31.6 {\rm kHz}$. They can be fitted by an exponential, $\tau=\tau_0\exp(-(H-H_{\rm SP})^2/H_0^2)$, for $H>H_{\rm SP}$.

The maximum in $\Gamma_{\rm int}$ can be understood as a result of enhanced relaxation time near the spinodal point. In the region $\omega\tau\ll 1$,  while $k_{\rm int}\simeq A(d)$ increases monotonically with decreasing $d$, $\Gamma_{\rm int}\simeq a A(d)\omega\tau_d$ is determined by the competition between $A(d)$ and $\tau_d$. When $\tau_d$ is singular, $\Gamma_{\rm int}$ is determined predominantly by $\tau_d$ near $H_{SP}$, and the maximum is located right at the spinodal point. When the singularity in $\tau_d$ is smeared, due to the distance dependence of tip-sample coupling, the maximum moves away from the spinodal point towards smaller $d$ (see Fig. \ref{fit4K}(d)(h)(l) for numerical fit).

Let us now estimate the order of magnitude for the friction coefficient. Friction is essentially determined by the following three factors: the tip-sample interaction energy $J$, the surface impurity density $n_{\rm imp}$, and the characteristic energy scale of the surface susceptibility $E_C\sim1/ \langle {\cal S}_d(\omega)\rangle$. Including spin degeneracy, one can write the friction term as $\omega_0\Gamma_{\rm int}\simeq s^2 S^2n_{\rm imp}J^2/E_C$. The prefactor $s^2S^2$ is about 1-10. Assuming there are about 5 impurities per square nanometer, to get the experimental value of $\omega_0\Gamma_{\rm int}$, which is about $100 N/m$ for $d \simeq 2nm$ \cite{Saitoh10}, one needs to have $J^2/E_C \simeq 10-100 eV$. If we take $E_c$ to be of order $k_BT$, where $T=4.2$ is the temperature at which the experiment is performed \cite{Saitoh10}, then $E_c\simeq 3\times 10^{-4} eV$, and the coupling is $J\simeq 3-30 meV$, which can be achieved. 


{\it Charge-charge interaction:}
For completeness, we also consider the possibility that friction arises from interactions in the charge channel, though this mechanism may not apply for \cite{Saitoh10} . We notice that, as was   shown in \cite{Persson05} , coupling between charge on the cantilever tip and ion vibrations on the sample surface can produce strong enhancement of NCF. This mechanism may be responsible for the observed NCF in \cite{Stipe01, Kisiel11}.

Essentially same logic outlined earlier applies with obvious substitution of Coulomb interactions for spin interactions. We assume there are some randomly distributed two-level fluctuators (TLF) on the sample surface, formed from localized electronic trap states \cite{Rogers85,Yu07}. The charge on the tip of the cantilever provides an external electric field, favoring one of the two states. The TLF are now governed by the Hamiltonian,  ${\cal H}_{surf}=-\sum_{i\neq j}J_{ij}Q_i Q_j-\sum_{i}V_i Q_i$, with charge $Q_i=0, 1$, the coupling $J_{ij}$ a random number, and $V_i$ the electric potential created by the tip charge.

The result is qualitatively the same as the case with spin-spin interactions, though the order of magnitude can be different. Here $\omega_0\Gamma_{\rm int}\simeq q^2n_{\rm imp}J_{(e)}^2/E^{(e)}_C$. For $d \simeq 2nm$, the Coulomb potential is $J_{(e)}\simeq 0.3eV$. Taking $E^{(e)}_C\simeq k_BT\simeq 3\times 10^{-4}eV$, to get the experimental result of $\omega_0\Gamma_{\rm int}\simeq 100 N/m$, one needs to have $q^2n_{\rm imp}\simeq 2$, with $q$ the tip charge in unit of the elementary charge $e$, and $n_{\rm imp}$ the number of surface charges per square nanometer. This can be easily achieved.

{\it Conclusion:} In conclusion, we have proposed a general mechanism to explain the distance dependence of the friction coefficient and the induced spring constant of an oscillating cantilever.  A universal ingredient of the proposed mechanism is the backaction effects of relaxational dynamics of the defects on the sample surface. This mechanism also explains nicely the observed order of magnitude of the friction coefficient. Furthermore, our formalism provides a general framework for experimentalists to extract separate information about tip-sample coupling and surface dynamics, thus enabling more detailed investigation of surface properties. One way to test our theory would be to examine the magnetic field dependence of the friction coefficient and the induced spring constant.

We acknowledge useful discussions with Gennady Berman, Tanmoy Das, Jason Haraldsen, Dima Mozyrsky, John Mydosh, David Sherrington and Keiya Shirahama. We are grateful to Kohta Saitoh, Kenichi Hayashi, Yoshiyuki Shibayama and Keiya Shirahama for allowing us to use their data. This work was supported, in part, by UCOP-TR01, by  the Center for Integrated Nanotechnologies, a U.S. Department of Energy, Office of Basic Energy Sciences user facility and in part by LDRD.  Los Alamos National Laboratory, an affirmative action equal opportunity employer, is operated by Los Alamos National Security, LLC, for the National Nuclear Security Administration of the U.S. Department of Energy under contract DE-AC52-06NA25396.

\section*{Appendix I: Divergence of relaxation time near the spinodal point}

\begin{figure}
\begin{centering}
\includegraphics[width=0.4\linewidth]{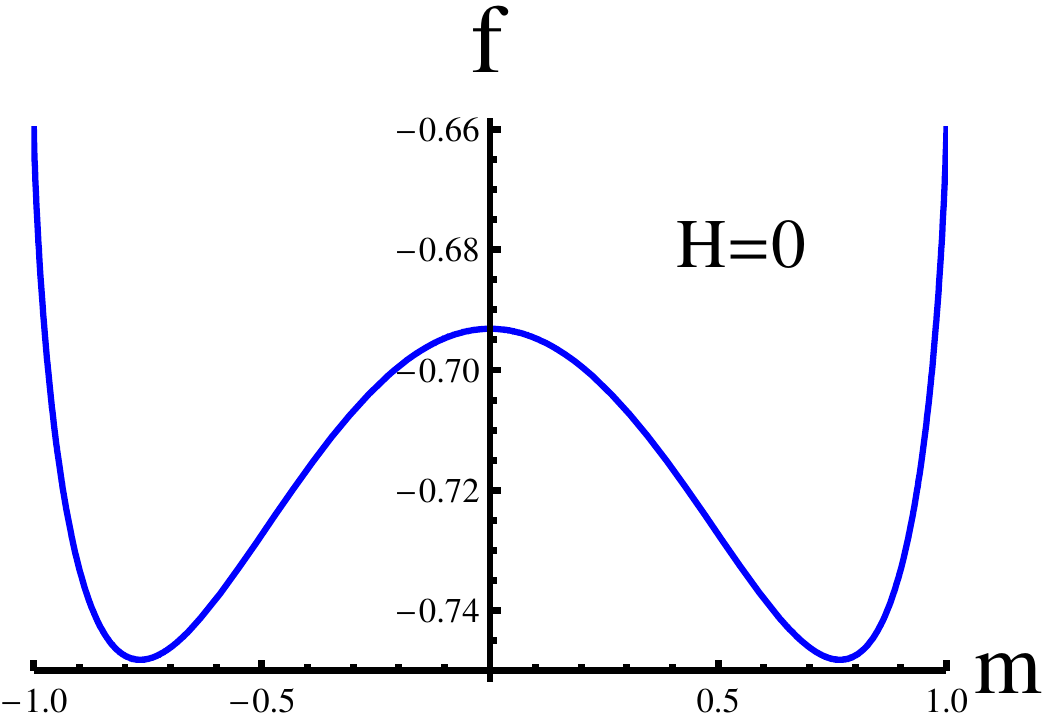} 
\includegraphics[width=0.4\linewidth]{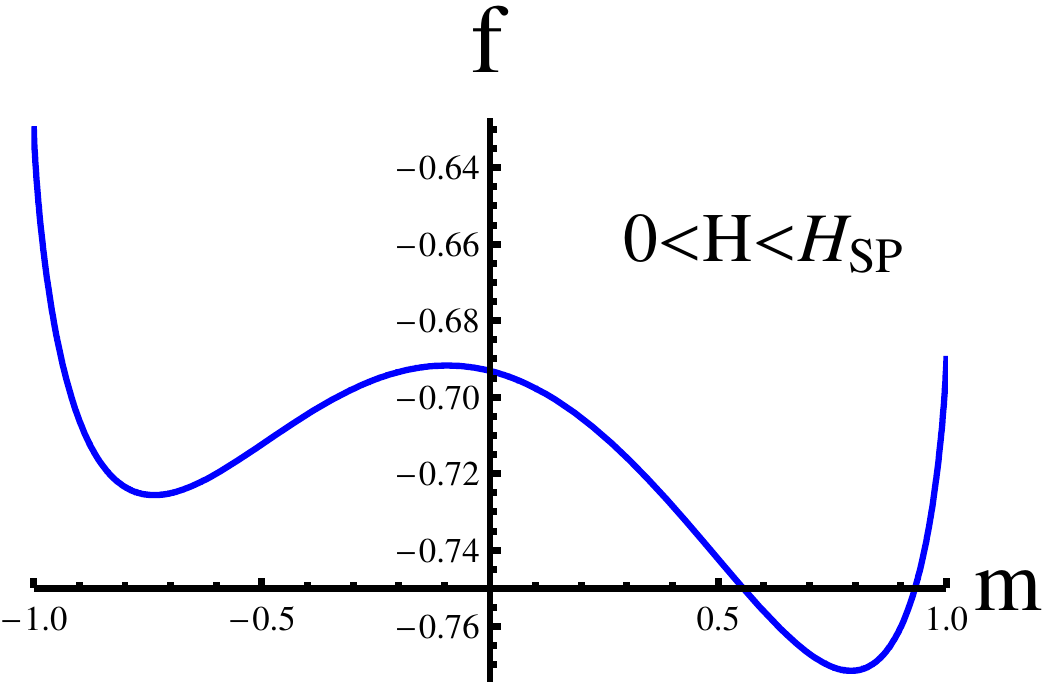} 
\includegraphics[width=0.4\linewidth]{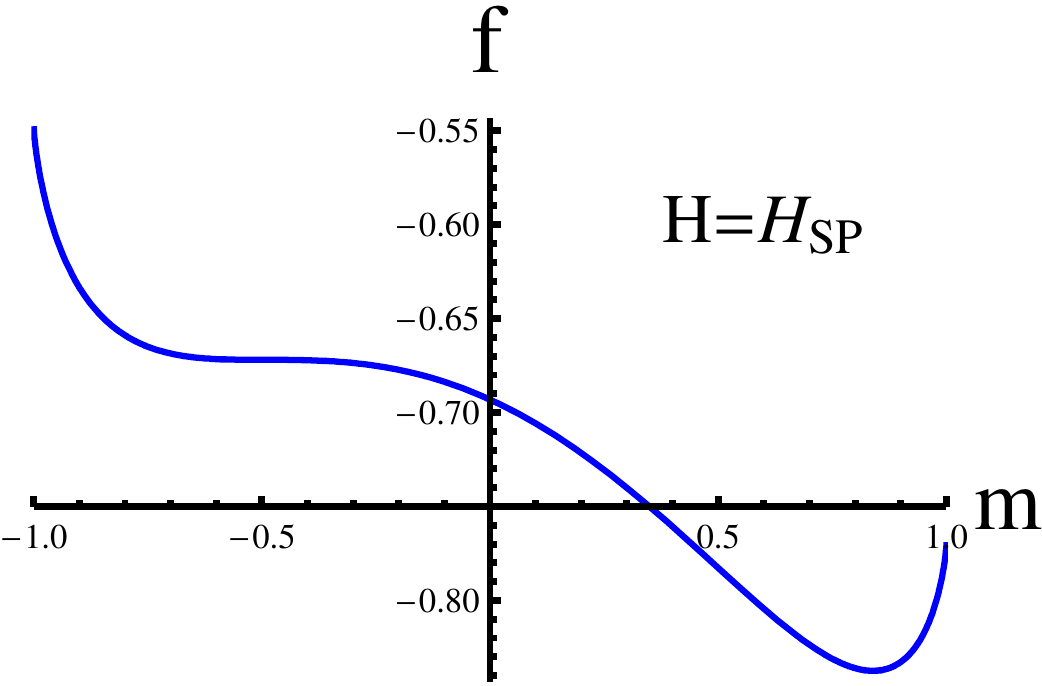} 
\includegraphics[width=0.4\linewidth]{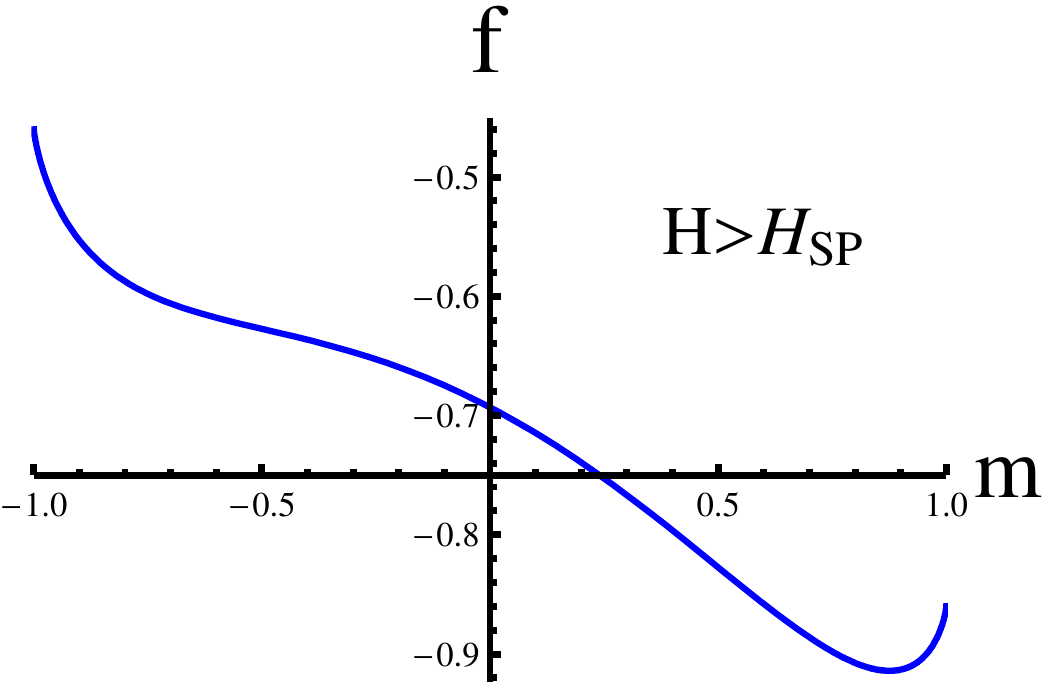} 
\end{centering}
\caption{Evolution of potential energy with magnetic field in Husimi-Temperley model. We are plotting $f(m)=-\frac{J}{2}m^2-Hm+\frac{1}{\beta}\left( \frac{1+m}{2}\ln  \frac{1+m}{2}+\frac{1-m}{2}\ln  \frac{1-m}{2} \right)$ \cite{Mori10}, with $\beta=1, J=1.32, H=0, 0.03, 0.11, 0.2$.}
\label{fSP}
\end{figure}

We consider here the long-range, weakly interacting Husimi-Temperley model where each spin $\sigma_i=\pm 1$ interacts equally with every other spin. In the presence of an external magnetic field $H$, the Hamiltonian reads
\begin{equation}
{\cal H}=-\frac{J}{2N}M^2-HM, ~~~~ M=\sum_{i=1}^N\sigma_i,
\end{equation}
When $H=0$, the free energy $f$ has two stable minima, and this model displays a second-order phase transition. In the presence of a weak magnetic field, one minimum becomes a metastable state. At some critical field strength, this metastable state becomes unstable. At this value of field strength, the potential changes from a double well structure to a single well structure, and the metastable minimum and the barrier combine to form a saddle point, known as the spinodal point (see Fig.\ref{fSP}). 
This point is determined by the condition: $\partial f/\partial m=0, \partial^2 f/\partial m^2=0$, with $m=M/N$ the magnetization per spin.  

 Near the spinodal point, the relaxation of the system slows down dramatically. For $H>H_{\rm SP}$, the relaxation time is of the form \cite{Mori10, Binder89}
\begin{equation}
\tau\sim \frac{1}{\left|H-H_{\rm SP}\right|^{1/2}}\exp\left(b \left|H-H_{\rm SP}\right|^{3/2} \right), 
\end{equation}
when $\Lambda\equiv \beta N^{2/3}\left|H-H_{\rm SP}\right|\gg 1$. 
Here $b=(4/3)\beta N/\sqrt{J(\beta J-1)}$. Consider the double scaling limit $N\to \infty, \left|H-H_{\rm SP}\right|\to 0$ with $\Lambda$ large but finite, the relaxation time reduces to $\tau\sim \left|H-H_{\rm SP}\right|^{-1/2}$. For  $H>H_{\rm SP}$, there is a critical divergence of the relaxation time, \cite{Binder73}
\begin{equation}
\tau\sim\left|H-H_{\rm SP}\right|^{-1/2}. 
\end{equation}

In models with shorter range interactions, the change in relaxation time is smeared.

\section*{Appendix II: Friction at room temperature}
\begin{figure}
\begin{centering}
\includegraphics[width=0.45\linewidth]{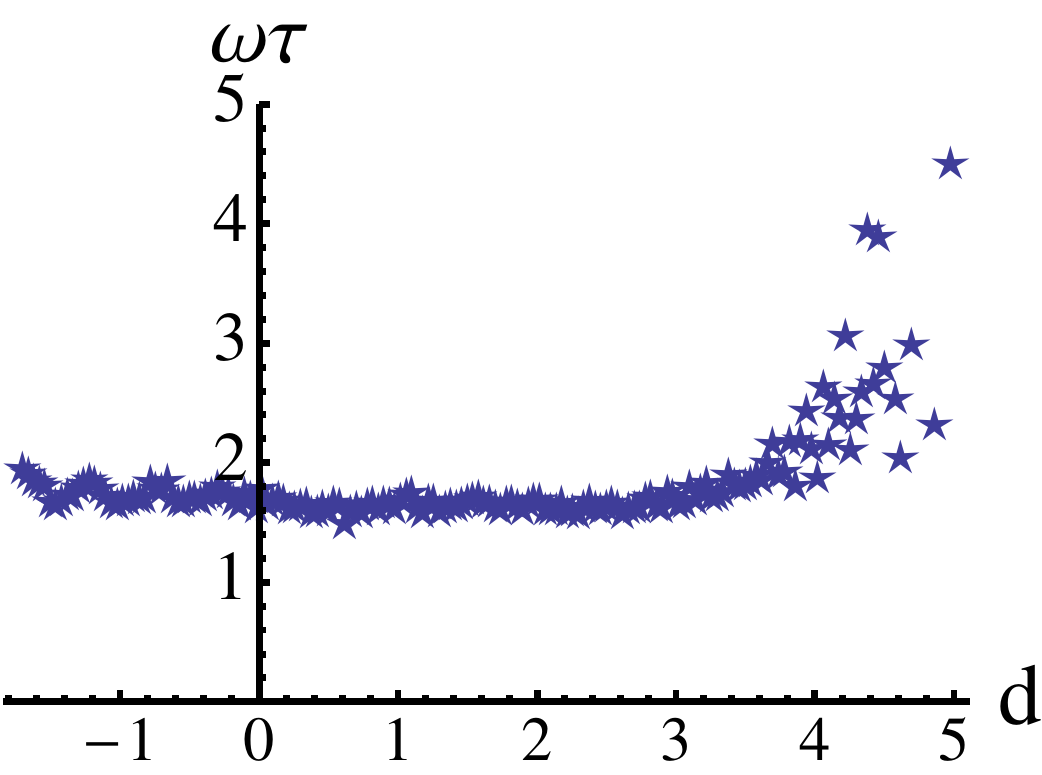} 
\includegraphics[width=0.45\linewidth]{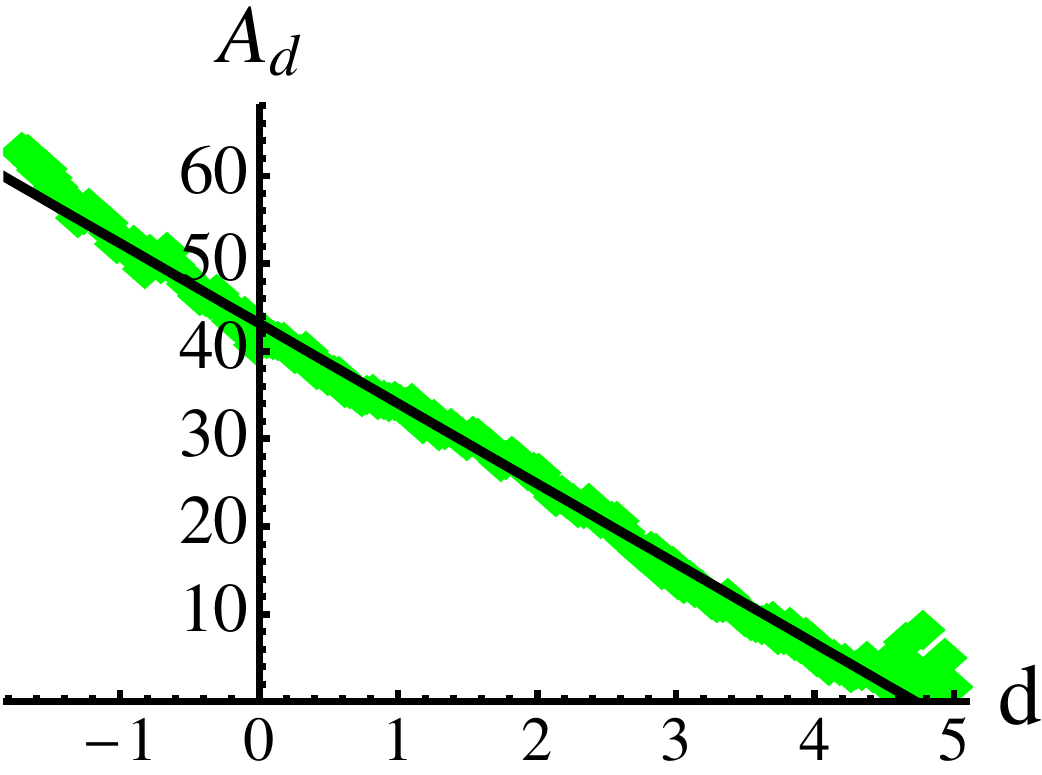}
\end{centering}
\caption{Relaxation time and tip-sample coupling for NbSe$_2$ at room temperature and $f_0=31.6 {\rm kHz}$. The star and diamond are experimental values (data by Saitoh et al. \cite{Saitoh10} ). The solid line is a fit to $A(d)=a_0+a_1d$ with $a_0=43.2, a_1=-9.1$.}
\label{wtAdHigh}
\end{figure}

Using the same method as in the main text, we extract here the relaxation time and tip-sample coupling at high temperature. The results are shown in Fig.\ref{wtAdHigh}. We can see that both of them display qualitatively different behavior as compared to the low temperature case. 
In the region where the experimental data can be trusted, i.e. $d\lesssim 4nm$, $\omega\tau_d$ is essentially distance independent; $A_d$ sets in at certain value of tip-sample distance $\sim 4.5 nm$, and then grows linearly in distance. These two features can be understood  by simply assuming that at high temperatures, there exists certain viscous cloud above the sample surface, extending up to $\sim 4.5 nm$ high. The NCF that's at work at low temperatures is now
suppressed by temperature effects, and the observed friction is due to the cantilever interacting with such cloud. Since it is always the same cloud, the relaxation time should not change with distance. Friction is proportional to the length of the part of the cantilever that is
immersed in this cloud. 
 
\bibliographystyle{apsrev}
\bibliography{strings,refs}

\end{document}